\documentclass[a4paper,11pt]{article}
\usepackage{pos}

\title{\texttt{ggxy}: Fast and flexible NLO QCD corrections to gluon-initiated processes}

\author*[a]{Daniel Stremmer}

\affiliation[a]{ Institut f{\"u}r Theoretische Teilchenphysik, Karlsruhe Institute of Technology (KIT), \\Wolfgang-Gaede Stra\ss{}e 1, 76131 Karlsruhe, Germany}

\emailAdd{daniel.stremmer@kit.edu}

\abstract{We present the \texttt{C++} library \texttt{ggxy}, which can be used for the fast and flexible calculation of partonic and hadronic cross sections to gluon-initiated top-mediated processes such as $gg\to HH$, as well as the newly derived processes $gg\to ZH$ and $gg\to ZZ$, at NLO QCD including full top-quark mass dependence. The two-loop virtual amplitudes are implemented using analytical approximations in different kinematic regions, while all other parts of the calculation are exact. This implementation allows to freely modify all input parameters, such as the top-quark mass and its renormalization scheme and the masses of the external $Z$/Higgs bosons. In addition, \texttt{ggxy} has been interfaced to Powheg, which allows the matching to parton showers.

\bigskip

P3H-26-049, \, TTP26-021 
}

\FullConference{Loops and Legs in Quantum Field Theory (LL2026)\\
12-17, April, 2026\\
Bayreuth, Germany\\}

\begin{document}
\maketitle

\section{Introduction}
The continuos improvements on the experimental and the greatly reduced statistical uncertainties expected from the High-Luminosity LHC (HL-LHC) requires also improvements on the theoretical side to reduce uncertainties to exploit the full potential of new measurements. This on the one hand requires the calculation of difficult higher-order corrections and on the other hand to make the numerical evaluation of them efficient and stable.

In these proceedings we discuss the calculation of the two-loop QCD corrections with full top-quark mass effects of $gg\to ZH$ and $gg\to Z^\star Z^\star$. The results of $gg\to ZH$ have been implemented to the public \texttt{C++} library \texttt{ggxy} \cite{Davies:2025qjr,Davies:2026uxl}  and provide a fast and stable numerical evaluation of the amplitudes and hadronic cross sections. This is achieved by the computation of the two-loop amplitudes in the high-energy and the forward limits that combined cover the whole phase space \cite{Davies:2025out}. In addition, we use these results to implement the $gg\to \ell^-\ell^+H$ and $gg\to \nu_\ell \bar{\nu}_\ell H$ process at NLO QCD preserving spin correlations and off-shell effects of the $Z$ decay in \texttt{POWHEG}~\cite{Alioli:2010xd}, which can be used for the combination with parton showers from e.g.~\texttt{Pythia}~\cite{Sjostrand:2007gs} or \texttt{Herwig}~\cite{Bellm:2025pcw}.

While our results implement in \texttt{ggxy} represents the first public available implementation of $gg\to ZH$ at NLO QCD with full top-quark mass dependence, the two-loop corrections have also been calculated using different methods. In particular, the loop-induced LO order contribution has been calculated in Refs.~\cite{Kniehl:1990iva,Dicus:1988yh,Dicus:1987dj,Glover:1988rg}. The NLO QCD corrections have been first calculated in the large-mass expansion \cite{Altenkamp:2012sx,Melnikov:2015laa,Hasselhuhn:2016rqt,Davies:2020drs,Campbell:2016ivq} followed by calculations in the high-energy expansion~\cite{Davies:2020drs,Davies:2020lpf}, the small-mass expansion~\cite{Wang:2021rxu} and the $p_T$ expansion~\cite{Alasfar:2021ppe,Degrassi:2024fye}. In contrast to these expansion methods, the two-loop amplitudes have also been computed in
Refs.~\cite{Chen:2020gae,Agarwal:2020dye,Chen:2022rua,Agarwal:2024pod} with numerical methods. The two-loop corrections to $gg\to ZZ$ induced by a light-quark loop were computed in Refs.~\cite{vonManteuffel:2015msa,Caola:2015ila}.

These proceedings are organized as follows: In Section \ref{sec1} we discuss the computation of the two-loop amplitudes of both processes using the high-energy and forward expansions. Numerical results of the two-loop amplitudes are presented in Section \ref{sec2}. In Section \ref{sec3} we describe the implementation of the $gg\to ZH$ process in \texttt{ggxy}.

\section{\label{sec1} Technical details}
The main bottleneck in the calculation of NLO QCD corrections with full top-quark mass dependence for $gg\to ZH$ and $gg\to Z^\star Z^\star$ is the computation of the two-loop amplitudes. This is circumvented by the computation of those amplitudes in the high-energy and the forward limits \cite{Davies:2025out} that not only cover together the whole phase space but also provide an efficient and numerically stable way to compute the two-loop amplitudes as already demonstrated for the simpler $gg\to HH$ process \cite{Davies:2025qjr}. The starting point of both expansions is the same that the amplitudes are expanded in the limit where the external masses of final-state particles, $q_3^2$ and $q_4^2$, are smaller than all other scales involved in the process. This expansion corresponds to a Taylor expansion so that the amplitudes can afterwards be written as a sum of master integrals $(M_i(s,t,m_t^2,\epsilon))$ that do not depend on the external masses. The master integrals are next expanded in the two different limits, where in the forward expansion we consider $|t|\ll s,m_t^2$ and in the high-energy expansion we have $m_t^2\ll s,|t|$. While the former expansion is again a Taylor expansion, in the high-energy limit we have to consider an asymptotic expansion. The results of the forward limit can alternatively be obtained by performing an expansion in the small variable $\delta q$ that is introduced in the amplitude with $q_3=-q_1+\delta q$, where $q_3$ is the momentum of one of the two final-state particles and $q_1$ of one of the two gluons. This method leads to a simultaneous expansion in the small variables $t$, $q_3^2$ and $q_4^2$ so that a deeper expansion in the external masses is possible but less terms in the $t$ expansion can be achieved compared to the first method. In addition, we construct a Pad\'{e} approximation for the amplitudes obtained in the high-energy expansion following Ref.~\cite{Davies:2020lpf} to increase the region of convergence.

\section{\label{sec2} Numerical results for two-loop amplitudes}

\begin{figure}[t]
    \begin{center}
    \begin{tabular}{cc}
    \includegraphics[width=.45\textwidth]{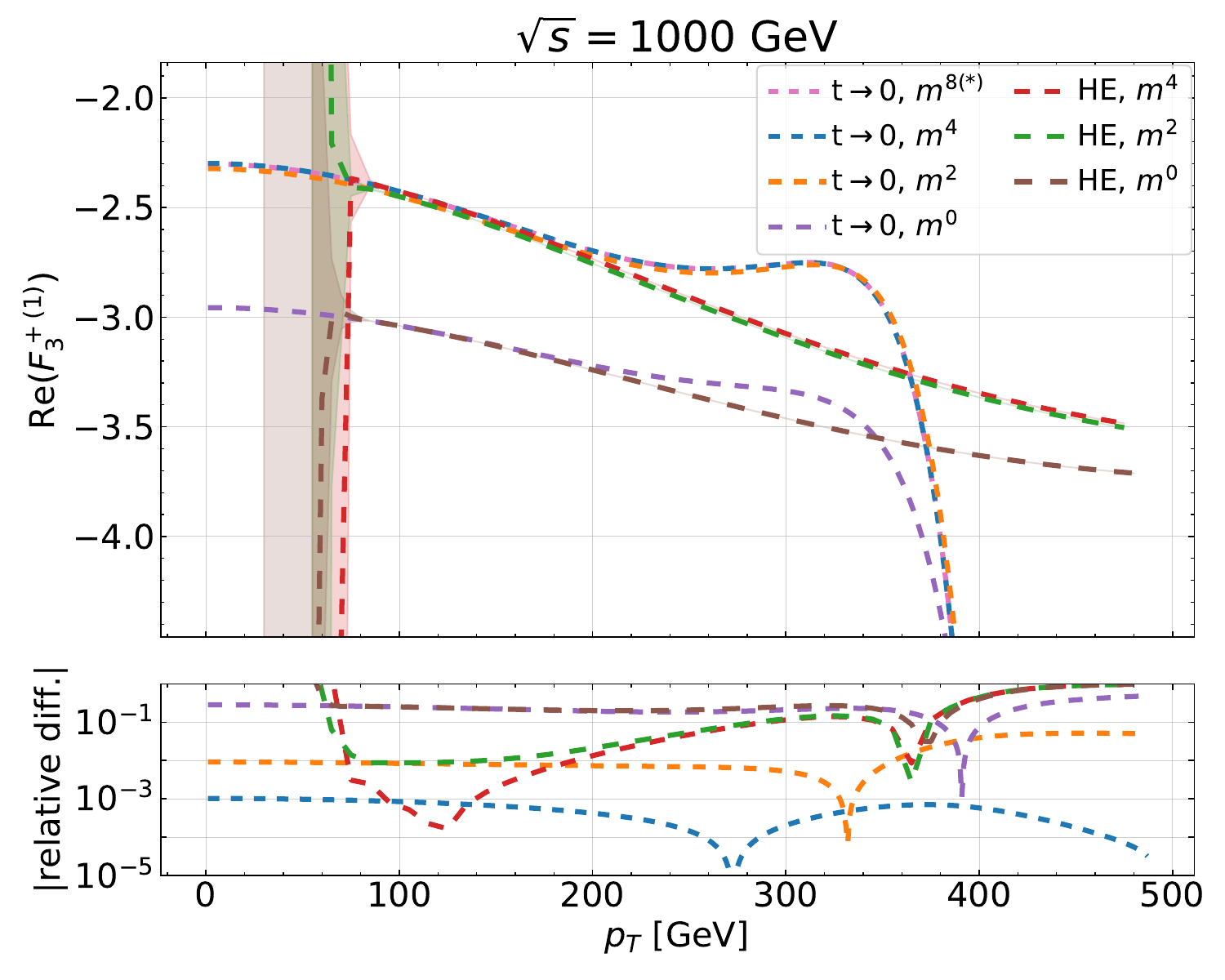} &
    \includegraphics[width=.45\textwidth]{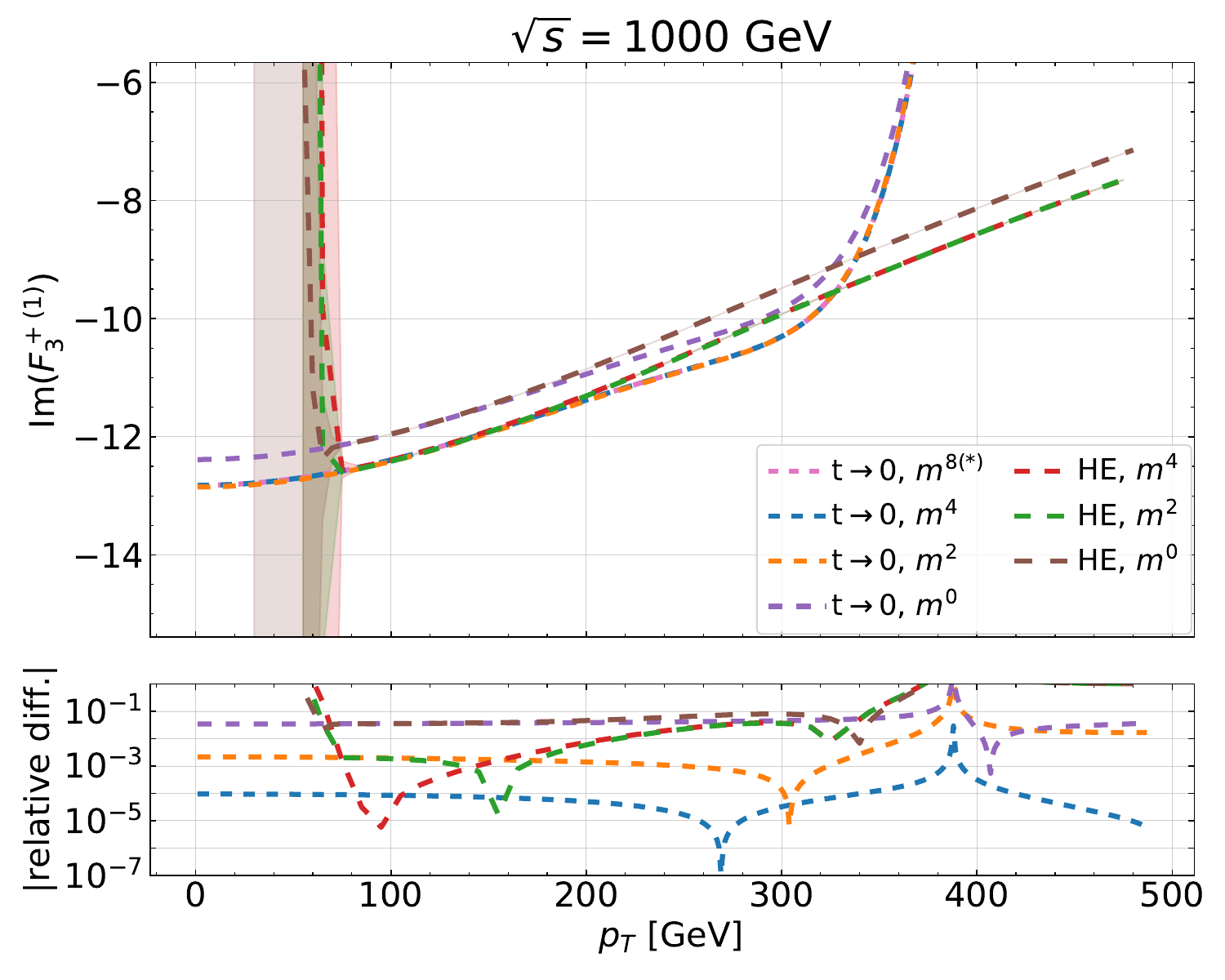}
    \end{tabular}
    \end{center}
  \vspace{-20pt}
  \caption{\label{fig::zh_nlo_fixsqrts}
  Real and imaginary parts of $F_3^{+(1)}$ for $gg\to ZH$ as a function of $p_T$ for $\sqrt{s}=1000$~GeV. High-energy and $t\to 0$ expansions are shown including mass corrections up to $m_{Z,H}^{\{0,2,4\}}$. Also shown are higher mass corrections up to $m_{Z,H}^8$ for the $t\to 0$ expansion with fewer expansion terms in $t$ according to $t^{n_t}(m_Z^2)^{n_3}(m_H^2)^{n_4}$ with $n_t+n_3+n_4\leq 4$ denoted by $\star$. Lower panels display the relative difference with respect to the best approximation of the $t\to 0$ expansion. Figure taken from Ref. \cite{Davies:2025out}.}
\end{figure}

\begin{figure}[t]
    \begin{center}
    \begin{tabular}{cc}
    \includegraphics[width=.45\textwidth]{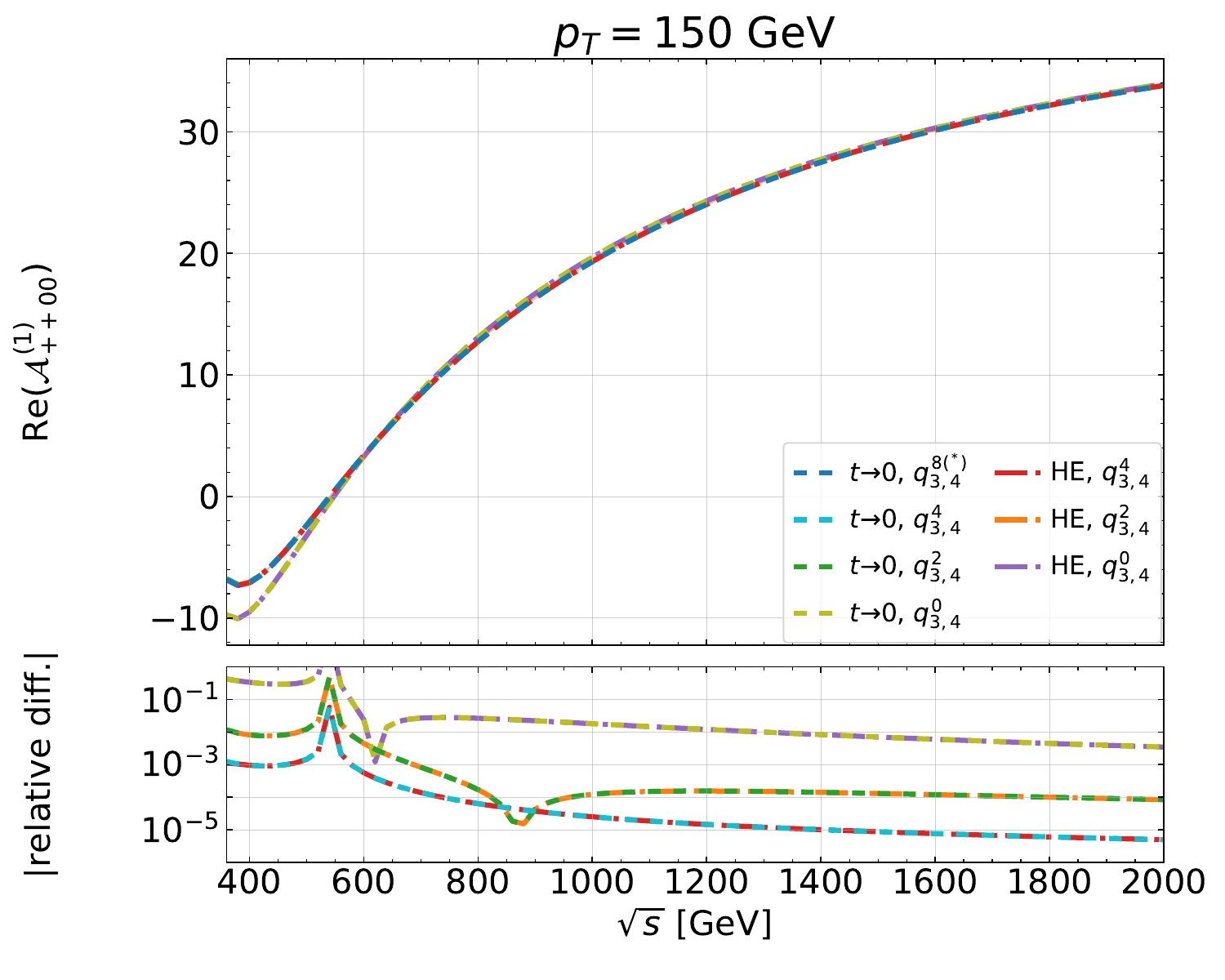} &
    \includegraphics[width=.45\textwidth]{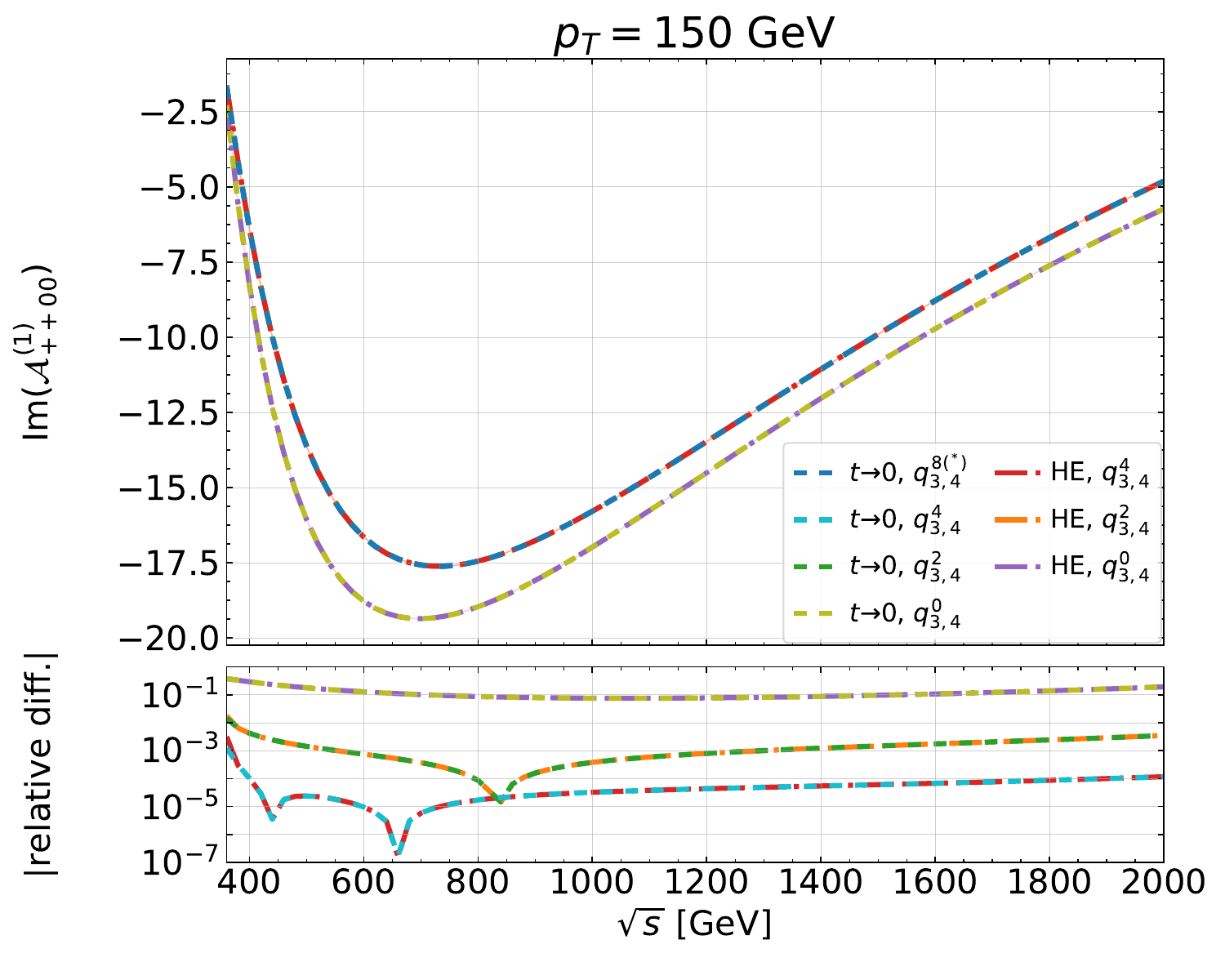}
    \end{tabular}
    \end{center}
  \vspace{-20pt}
  \caption{\label{fig::ggzz2l_pt150}
  Same as in Figure \ref{fig::zh_nlo_fixsqrts} for the helicity amplitude ${\cal A}^{(1)}_{++++}$ for $gg\to Z^\star Z^\star$ as a function of $\sqrt{s}$ for $p_T=150$~GeV. Figure taken from Ref. \cite{Davies:2025out}.}
\end{figure}

In order to demonstrate the general convergence pattern of both expansion methods, we show in Figure \ref{fig::zh_nlo_fixsqrts} the real and imaginary parts of the form factor $F_3^{+(1)}$ for the $gg\to ZH$ process as function of $p_T$ for $\sqrt{s}=1000$~GeV. Results are shown for the high-energy and forward ($t\to 0$) expansions including different number of terms in the mass expansion. We find that both expansions agree well with each other typically in the phase-space region around $p_T=200$~GeV, where both methods can be safely used for precise results. 
Towards smaller $p_T$ values only the $t\to 0$ expansion leads to reliable results while the high-energy expansion diverges, For high $p_T$, the behaviour is reversed, and only the high-energy expansion is accurate. In Figure \ref{fig::zh_nlo_fixsqrts} we show the real and imaginary parts of the helicity amplitude ${\cal A}^{(1)}_{++++}$ for the $gg\to Z^\star Z^\star$ process as a function of $\sqrt{s}$ for $p_T=150$~GeV. We find that both expansion methods lead to essentially the same results for all possible values of $\sqrt{s}$. Therefore, the region of convergence of both methods is limited only by the value of $p_T$, which is then used to determine which expansion method should be used. Note that the convergence of the high-energy expansion is additionally limited to the region above the top-quark pair threshold, $\sqrt{s}>2m_t\sim 345$~GeV. In practise, such small values are never used for the high-energy expansion due to the lower limit on $p_T$.

\section{\label{sec3} Hadronic cross sections with \texttt{ggxy}}
We have implemented the results for $gg\to ZH$ into our \texttt{C++} library \texttt{ggxy}\footnote{\textrm{\url{https://gitlab.com/ggxy/ggxy-release}}} which can be used for the calculation of the one- and two-loop amplitudes as well as the hadronic cross section at NLO QCD. Furthermore, \texttt{ggxy} is also used to provide a process implementation\footnote{\textrm{\url{https://gitlab.com/POWHEG-BOX/V2/User-Processes/ggxy_ggZH}}} of $gg\to\ell^-\ell^+H$ and $gg\to \nu_\ell\bar{\nu}_\ell H$ in \texttt{POWHEG}, which makes the matching to parton showers possible.

The one-loop $2\to 3$ amplitudes required for the real corrections are computed with the tools \texttt{Recola}~\cite{Actis:2016mpe}, \texttt{Collier}~\cite{Denner:2016kdg},  \texttt{CutTools}~\cite{Ossola:2007ax} and \texttt{OneLOop}~\cite{vanHameren:2010cp}. The calculation of the real corrections is organized with the Catani-Seymour dipole subtraction scheme \cite{Catani:1996vz} and the phase-space integration is performed with \texttt{avhlib}~\cite{vanHameren:2007pt,vanHameren:2010gg}. In addition, the conversion of the on-shell top-quark mass to the one in the $\overline{\rm MS}$ scheme and the running of it is obtained with \texttt{CRunDec}~\cite{Herren:2017osy}. For the numerical evaluation of polylogarithmic functions we use the code of Ref. \cite{Frellesvig:2016ske}.

\begin{table}[t]
    \centering
    \renewcommand{\arraystretch}{1.2}
    \begin{tabular}{cc@{\hskip 10mm}l@{\hskip 10mm}l@{\hskip 10mm}}
        \hline
        $\sqrt{s}$&
        &{ \tt ggxy}&Ref.~\cite{CampilloAveleira:2025rbh}  \\
        \hline
        $13$~TeV & $\sigma^{\rm LO}$ [fb]& $63.97(2)^{+26.7\%}_{-20.0\%}$ &    $64.0^{+27\%}_{-20\%}$\\
        & $\sigma^{\rm NLO}$ [fb]& $118.40(8)^{+16.5\%}_{-14.0\%}$ & $118.4^{+17\%}_{-14\%}$\\
        \noalign{\smallskip}\hline\noalign{\smallskip}
        $13.6$~TeV & $\sigma^{\rm LO}$ [fb]& $70.59(2)^{+26.3\%}_{-19.7\%}$ &    $70.6^{+26\%}_{-20\%}$\\
        & $\sigma^{\rm NLO}$ [fb]& $130.50(9)^{+16.3\%}_{-13.8\%}$ & $130.5^{+16\%}_{-14\%}$\\

        \noalign{\smallskip}\hline\noalign{\smallskip}
        $14$~TeV & $\sigma^{\rm LO}$ [fb]& $75.15(2)^{+26.0\%}_{-19.6\%}$ &    $75.2^{+26\%}_{-20\%}$\\
        & $\sigma^{\rm NLO}$ [fb]& $139.0(1)^{+16.3\%}_{-13.7\%}$ & $138.9^{+16\%}_{-14\%}$\\
        \noalign{\smallskip}\hline\noalign{\smallskip}
    \end{tabular}
    \caption{\label{tab::sig2} Comparison with results of Ref.~\cite{CampilloAveleira:2025rbh}
    at $\sqrt{s}=\{13,13.6,14\}$~TeV. Tables taken from Ref. \cite{Davies:2026uxl}.}
\end{table}
We have performed several cross-checks at the integrated and differential level with existing results to ensure the correctness of our implementation of $gg\to ZH$ at NLO QCD. An example of these cross-checks is shown in Table \ref{tab::sig2}, where we compare the integrated cross section at LO and NLO QCD for several center-of-mass energies with the results of Ref.~\cite{CampilloAveleira:2025rbh}. Both computations agree well within in the statistical uncertainties. A runtime of about $90$ minutes is required to obtain a precision of $0.1\%-0.2\%$ on a single core, which makes the computation of $gg\to ZH$ roughly three times slower than the one of $gg\to HH$.

\begin{figure}[t]
  \begin{center}
  \begin{tabular}{cc}
   \includegraphics[width=.45\textwidth]{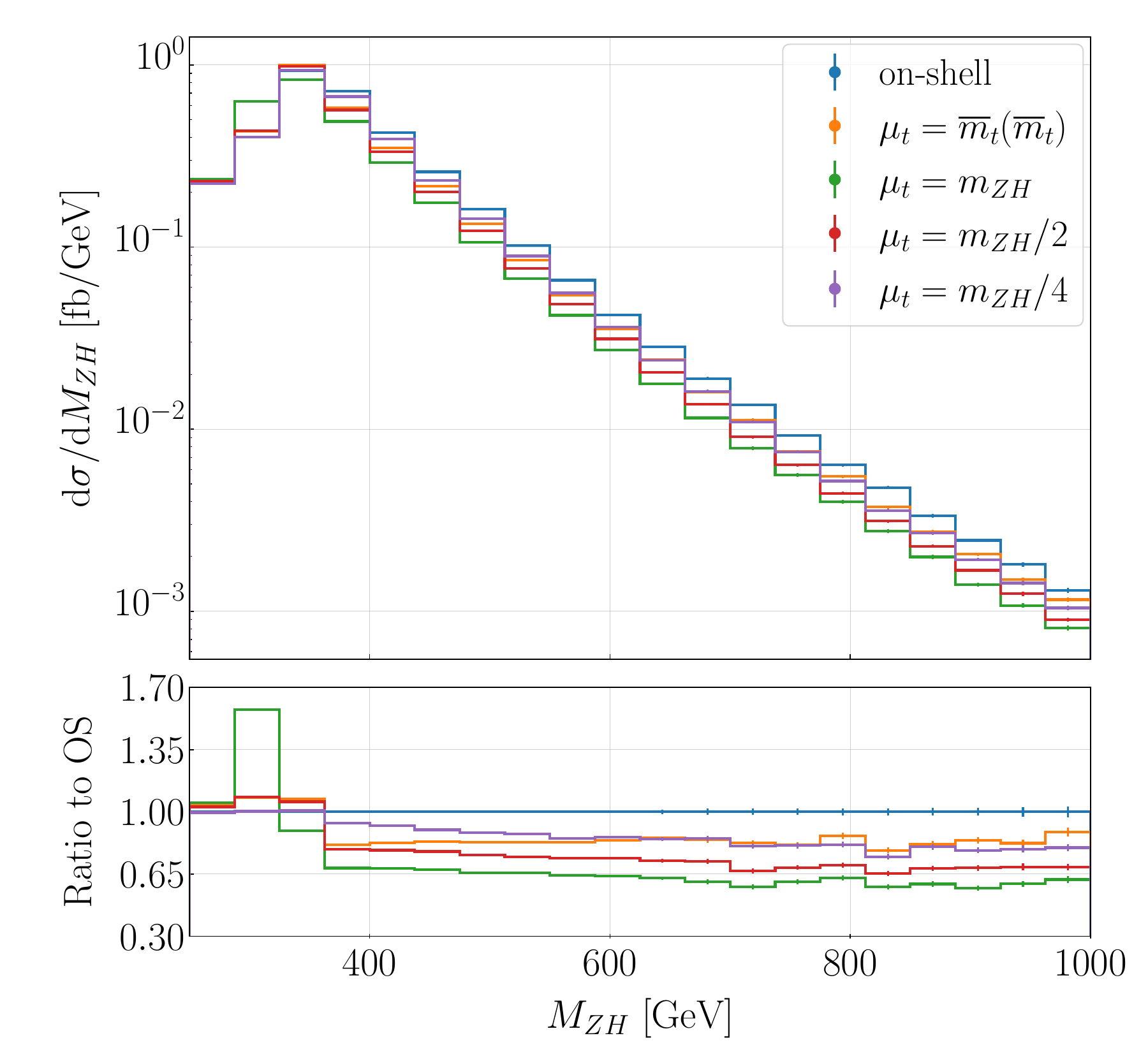} &
   \includegraphics[width=.45\textwidth]{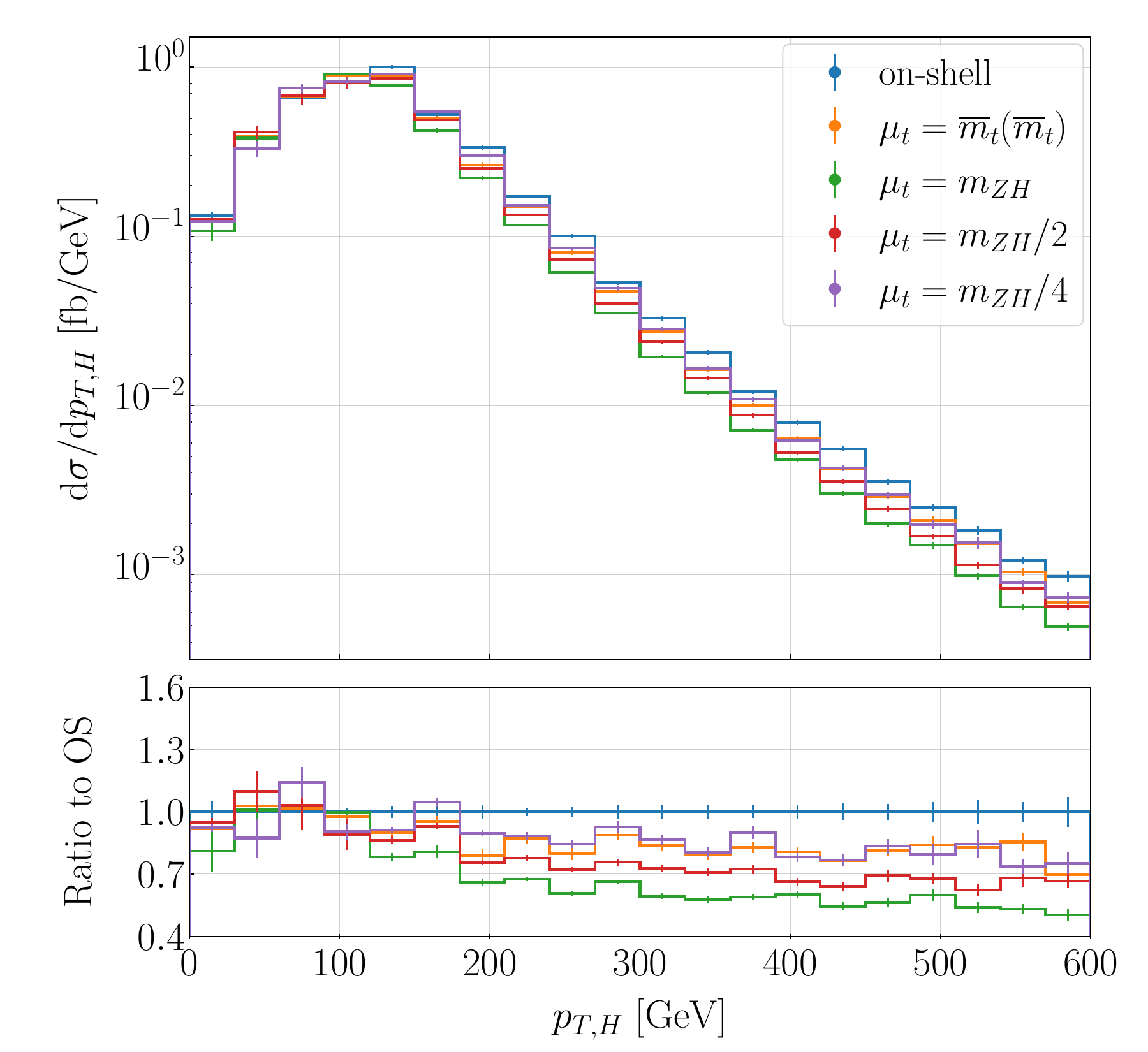}
  \end{tabular}
  \end{center}
  \vspace{-20pt}
  \caption{\label{fig::OSvsMS}
  Dependence on the top quark mass renormalization scheme for the distributions $M_{ZH}$ and $p_{T,H}$. Results are shown in the on-shell and $\overline{\rm{MS}}$ top-quark mass renormalization schemes for different values of the top-quark mass renormalization scale $\mu_t$. Lower panels show the ratio to the on-shell scheme. Figures taken from Ref. \cite{Davies:2026uxl}.}
\end{figure}
In Figure \ref{fig::OSvsMS} we show a comparison for the distributions $M_{ZH}$ and $p_{T,H}$ between the on-shell and $\overline{\rm MS}$ top-quark mass renormalization schemes at NLO QCD, where in the latter cases we use different values for the top-quark mass renormalization scale $\mu_t$. We find differences between these predictions up to $30\%-40\%$, which are similar size as the top-quark mass scheme differences for $gg\to HH$, see e.g. Ref. \cite{Davies:2025qjr}.
\begin{figure}[t]
  \begin{center}
  \begin{tabular}{cc}
   \includegraphics[width=.45\textwidth]{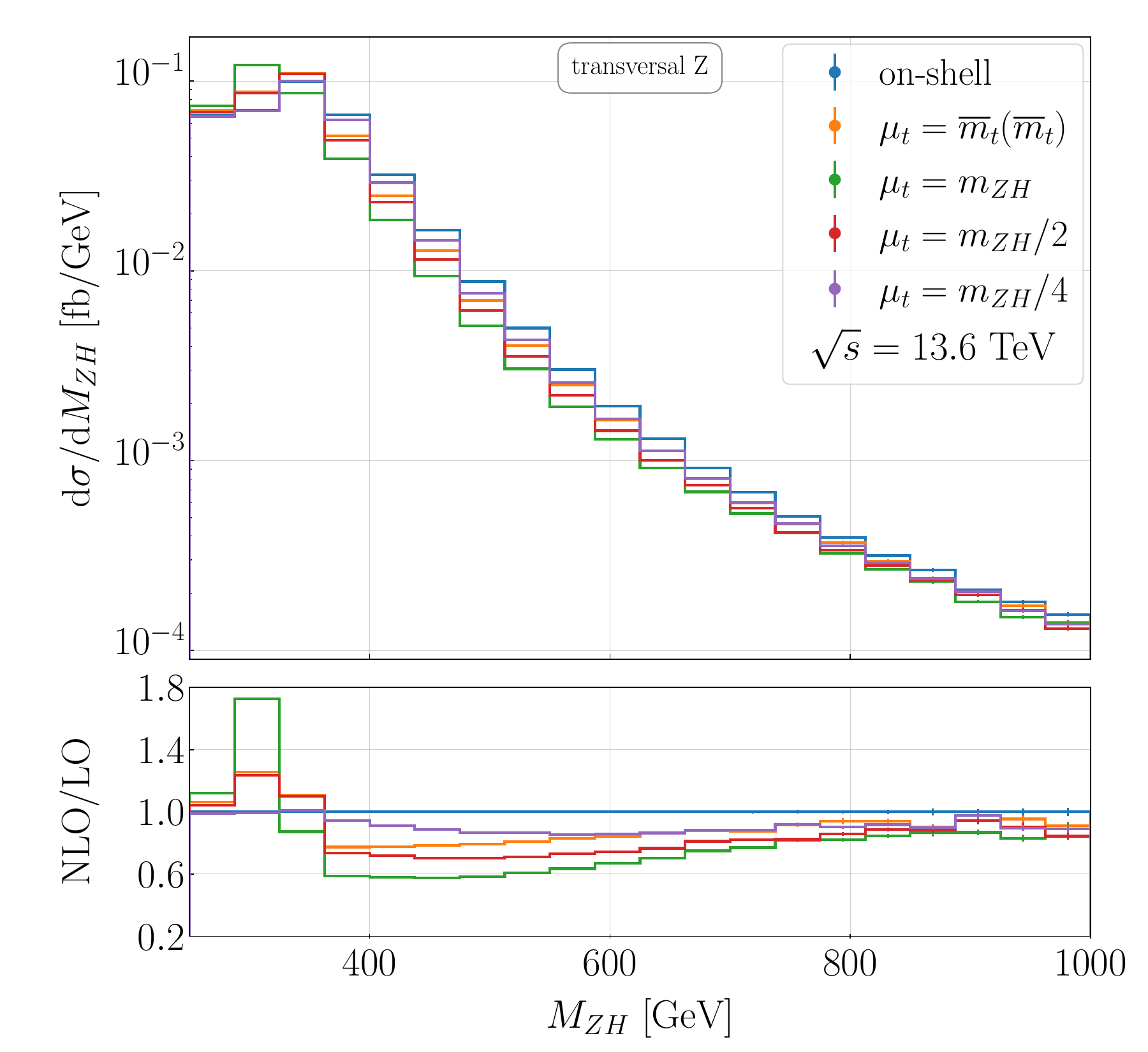} &
   \includegraphics[width=.45\textwidth]{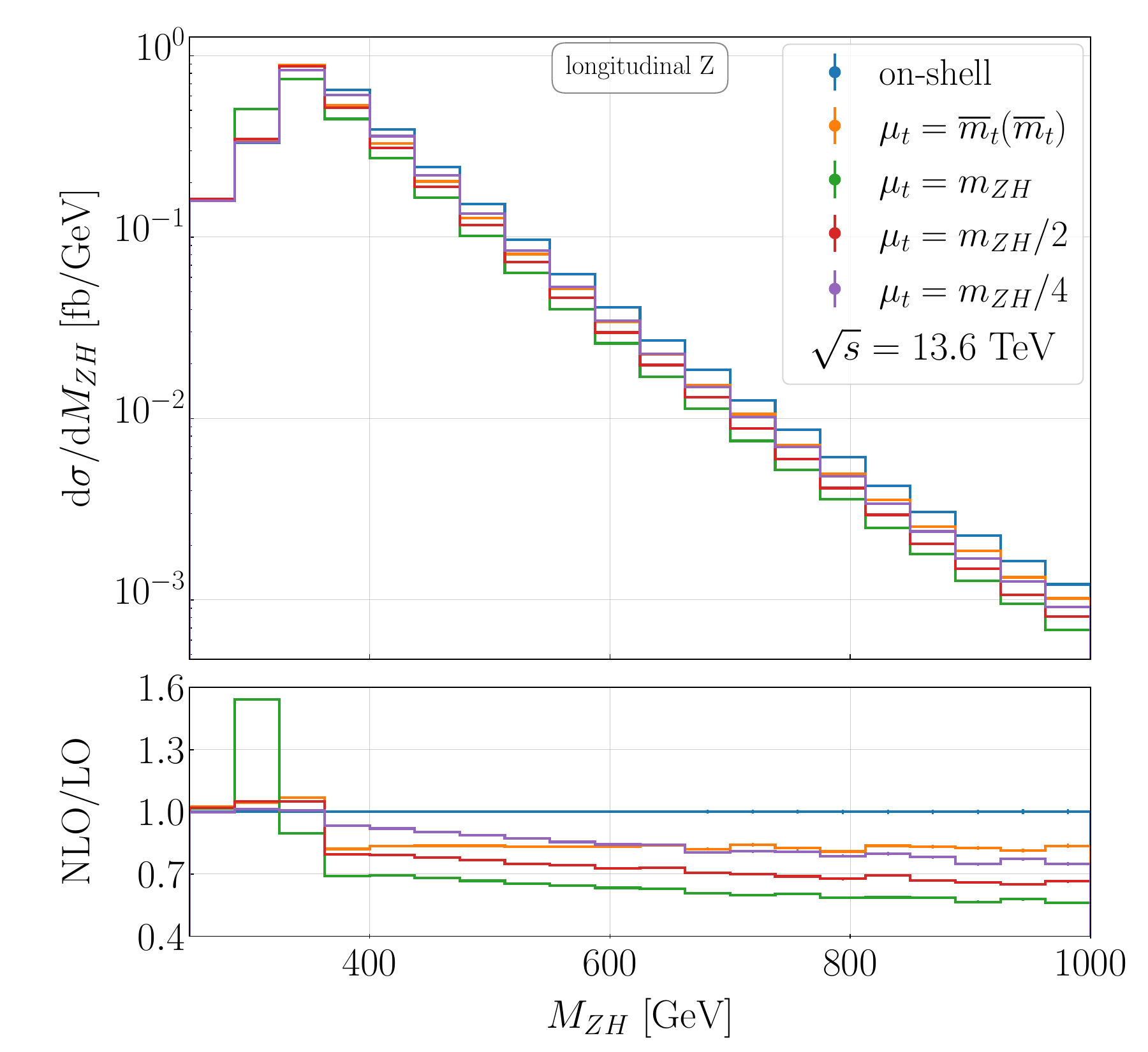}
  \end{tabular}
  \end{center}
  \vspace{-20pt}
  \caption{\label{fig::OSvsMS2}
  Same as Figure \ref{fig::OSvsMS} but for a transversely and longitudinally polarized $Z$ boson for the $M_{ZH}$ distribution.}
\end{figure}
To better understand the origin of these differences, we also study the top-quark mass scheme dependence for a polarized $Z$ boson. In Figure \ref{fig::OSvsMS2} we show the same comparison for a transversely and longitudinally polarized $Z$ boson for the $M_{ZH}$ distribution. We find similar large differences between the different predictions for the longitudinal polarization of the $Z$ boson, while the differences in the case of a transversely polarized $Z$ boson decrease towards the high-energy region. In addition, the contribution from the transversal polarization is suppressed by almost an order of magnitude, so that the effects for the unpolarized cross section is mainly coming from the longituidinal polarization of the $Z$ boson.

\section{\label{sec4} Conclusions and outlook}
In these proceedings we have presented the computation of the two-loop amplitudes for $gg\to ZH$ and $gg\to Z^\star Z^\star$ with full top-quark mass dependence using two different expansion methods that combined cover the full phase-space. The latter results can directly be reused for the $gg\to Z^\star\gamma^\star$ and $gg\to\gamma^\star\gamma^\star$ processes. The results of $gg\to ZH$ have been implemented in the \texttt{C++} library \texttt{ggxy} that can be used for the efficient and numerically stable evaluation of the two-loop amplitudes as well as the computation of the hadronic cross sections at NLO QCD. In addition to unpolarized hadronic cross sections, we have implemented the capability to compute hadronic cross section with a transversely or longitudinally polarized $Z$ boson. Furthermore, these results have been used for the implementation of $gg\to\ell^-\ell^+H$ and $gg\to \nu_\ell\bar{\nu}_\ell H$ at NLO QCD in \texttt{POWHEG}.

The extension of polarized $gg\to ZH$ production as well as the implementation of the amplitudes for $gg\to VV$ production will be available in a future version of \texttt{ggxy}.

\section*{Acknowledgments}
We would like to thank Kirill Melnikov for the suggestion to study the top-quark mass scheme dependence for polarized $Z$ bosons.
This research was supported by the Deutsche Forschungsgemeinschaft (DFG, German
Research Foundation) under grant 396021762 — TRR 257: {\it P3H - Particle Physics Phenomenology after the Higgs Discovery}.


\bibliographystyle{JHEP}
\bibliography{References.bib}

\end{document}